\documentclass{iau} 
\usepackage{graphicx}

\title[Dusty star-formation in the early Universe] 
{Exploring the dusty star-formation in the early Universe using intensity mapping}

\author[Guilaine Lagache]  
{Guilaine Lagache$^1$}

\affiliation{$^1$Aix Marseille Univ, CNRS, LAM, Laboratoire d'Astrophysique de Marseille \\ Marseille, France \\ email: {\tt guilaine.lagache@lam.fr}}
\pubyear{2018}
\volume{333}  
\jname{Peering towards Cosmic Dawn}
\editors{Vibor Jeli\'c \& Thijs van der Hulst, eds.}
\begin{document}

\maketitle

\begin{abstract}
In the last decade, it has become clear that the dust-enshrouded star formation contributes significantly to early galaxy evolution.
Detection of dust is therefore essential in determining the properties of galaxies in the high-redshift universe. This requires observations at the (sub-)millimeter wavelengths.
Unfortunately, sensitivity and background confusion of single dish observations on the one hand, and mapping efficiency of interferometers on the other hand, pose unique challenges to observers. One promising route to overcome these difficulties is intensity mapping of fluctuations which exploits the confusion-limited regime and measures the collective light emission from all sources, including unresolved faint galaxies. We discuss in this contribution how 2D and 3D intensity mapping can measure the dusty star formation at high redshift, through the Cosmic Infrared Background (2D) and [CII] fine structure transition (3D) anisotropies.
\end{abstract}

\firstsection 

\section{Introduction}
The cosmic history of star formation is one of the most fundamental observables in astrophysical cosmology.
The star formation rate (SFR) is the key driver of structure evolution in the interstellar medium of galaxies, and strongly influences galaxy formation and evolution. It is now well established that the SFR density peaks at  z$\sim$2, and declines exponentially at lower
redshift. The Universe was much more active in forming stars in the past with an SFR density about ten times higher
at  z$\sim$2 than is seen today \cite[(e.g. Madau \& Dickinson 2014)]{madau14}. It is also well known that the obscured star-formation is dominating the overall budget with half of the energy produced since the surface of last scattering absorbed and reemitted by dust (Dole et al. 2006).\\
Although there are evidences that the SFR density increases steadily from z = 8 to z$\simeq$2, our direct knowledge of dust-obscured star formation at these redshifts is, for the most part, limited to the rarest and most ultraluminous dusty galaxies. This leaves considerable uncertainty about how much SFR density may be missing in the UV census of that early phase of galaxy evolution. Since the discovery of the CIB (Puget et al. 1996) and first measurements of obscured SFR density (e.g. Gispert et al. 2000), the progress on measuring the SFR of dusty star forming galaxies at high redshift have been quite slow due to heightened observational challenges.
Observing the dusty star-formation at very high redshift and its distribution in the cosmic web requires
undoubtedly (sub-)mm experiments (see the extreme case of HDF 850.1 which remains invisible in the deepest optical images, Walter et al. 2012). On the one hand, single-dish observations
with large field of view, but with poor angular resolution, are limited to relatively bright galaxies due to
confusion. On the other hand, (sub-)mm interferometers, with their high angular resolution but small field of view,
are sensitive to faint galaxies but very limited in sky coverage.
Hence, the contribution of dusty star-forming galaxies to the global budget of star formation is still a matter of debate for $z\gtrsim3$.
An original approach to the problem is to study the large-scale spatial anisotropies of the collective emission from all the dusty star-forming galaxies emitting in one frequency band (2D) or in some convenient spectral lines (3D) in the confusion-limited experiments. 

\section{2D intensity mapping: Cosmic Infrared Background anisotropies}
\begin{figure}[!h]
\begin{center}
\includegraphics[width=3.4in]{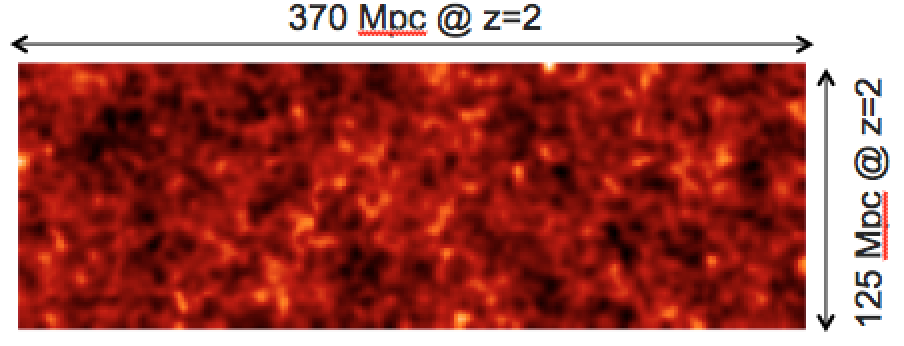} 
 \caption{CIB anisotropies as observed by Planck in the Bootes fields (the image size is $12^o\times4^o$). {\it Figure adapted from Planck collaboration (2014b)}.}
   \label{cib-bootes}
\end{center}
\end{figure}

The high redshift signal is embedded into the cosmic infrared background (CIB) anisotropies, that encode the
distribution of the emission of all dusty galaxies throughout cosmic history.  CIB anisotropies are quite bright ($\delta I/I \sim$15\%, Planck collaboration 2011). 
Due to a distinct frequency-redshift dependance (from 3000 GHz to 100\,GHz) they probe a large span of redshifts (Lagache et al. 2005) and they are an important tracer of large-scale structures (e.g. Planck collaboration 2014a). CIB anisotropies have been measured with a high signal to noise ratio with Herschel and Planck (Viero et al. 2013, Planck collaboration 2014b, see Fig.\,\ref{cib-bootes}). These key measurements have allowed to probe the clustering properties of dusty, star-forming galaxies and thus constrain the relationship between star formation and dark matter distribution, to measure the cosmic abundance of dust and the star formation rate density, and to derive the mean spectral energy distributions of high-redshift dusty galaxies. Deriving these quantities requires to model the angular power spectrum of CIB.

Making various assumptions such as the form of the relationship between galaxy luminosity and dark-matter halo mass, the typical halo mass for galaxies that dominate the CIB power spectrum has been found to lie between $\sim$1.3 and 6$\times$10$^{12}$\,M$\odot$ (Viero et al. 2013, Planck collaboration 2014b, Maniyar et al. 2018) and to be relatively constant from $z=4$ to $z=1$ (Maniyar et al. 2018). It is also shown that cosmic abundance of dust, relative to the critical density, is constant (within the error bars) at a level of $\sim$2-8$\times$10$^{-6}$ in the redshift range 1-3 (Thacker et al. 2013, Schmidt et al.2015). The fraction of dust mass compared to stellar mass is evolving from z$\sim$2 to z=0, from $\Omega_{dust} \sim 1\% \Omega_{M\star}$ to  $\Omega_{dust} \sim 0.2\% \Omega_{M\star}$.  The dust-to-stellar mass ratio depends strongly on the star formation history in galaxies. Calura et al. (2017) show that spiral galaxies have a nearly constant dust-to-stellar mass ratio as a function of the stellar mass and cosmic time, whereas proto-spheroid starburst galaxies present an early steep increase of the dust-to-stellar mass ratio, which stops at a maximal value and decreases in the latest stages.  The differential cosmic evolution of these galaxy populations is responsible for the variation of the dust-to-stellar mass ratio with redshift. \\

Interpretation of CIB measurements at $z\gtrsim3$ is currently limited due to the unknown redshift distribution of the underlying sources. For example, the degeneracy between redshifts and luminosities leads to very different histories of the SFR
density, depending on the modelling (see Planck collaboration 2014b). One can improve the situation by using all constraints together (e.g., adding dusty galaxy number counts and spectral energy distributions of galaxies to the measurements of CIB anisotropies --  e.g. B\'ethermin et al. 2013) or by putting strong priors (Maniyar et al. 2018). But the model uncertainties, combined with uncertainties in the observations at high $z$, make it impossible to isolate accurately the high-redshift signal ($z>4$) in the CIB. Dusty star formation has been found up to z$\sim$8 (Watson et al. 2015, Laporte et al. 2017). It is thus of primordial importance to quantify the obscured star formation at high redshift, at the key period when the interstellar medium of galaxies matures from a nearly primordial, dust-free state at $z\sim$8-10 to the dust- and metallicity-enriched state observed at $z\sim4$. Intensity mapping of a bright line (thus with a known redshift) is one of the most promising route (Kovetz et al. 2017).

\section{3D intensity mapping}

\par\bigskip\noindent
\subsection{Line intensity mapping}
In contrast to traditional large-scale structure surveys, intensity mapping measures the integrated light emission from all sources, including
unresolved faint galaxies. It statistically measures the properties of the light tracers and the underlying matter
distribution. Intensity mapping of a line retains redshift information (contrary to 2D intensity mapping).\\
In the last few years, considerable attention has been paid to intensity mapping using the 21\,cm
emission line of neutral hydrogen, as a powerful way of studying early structure formation, from the dark
ages, all the way through to the end of reionization. Indeed, measuring the fluctuations of the redshifted HI
line is one of the key science goals of LOFAR, the Square Kilometer Array and its pathfinders. These
experiments will trace the reionization of the intergalactic medium, but they will not observe the young stars responsible for it. To map the star formation at redshift $4.5<z<8.5$ an alternative is to use intensity mapping of the [CII] 157.7 micron line.

\begin{figure}[]
\begin{center}
\includegraphics[scale=0.32]{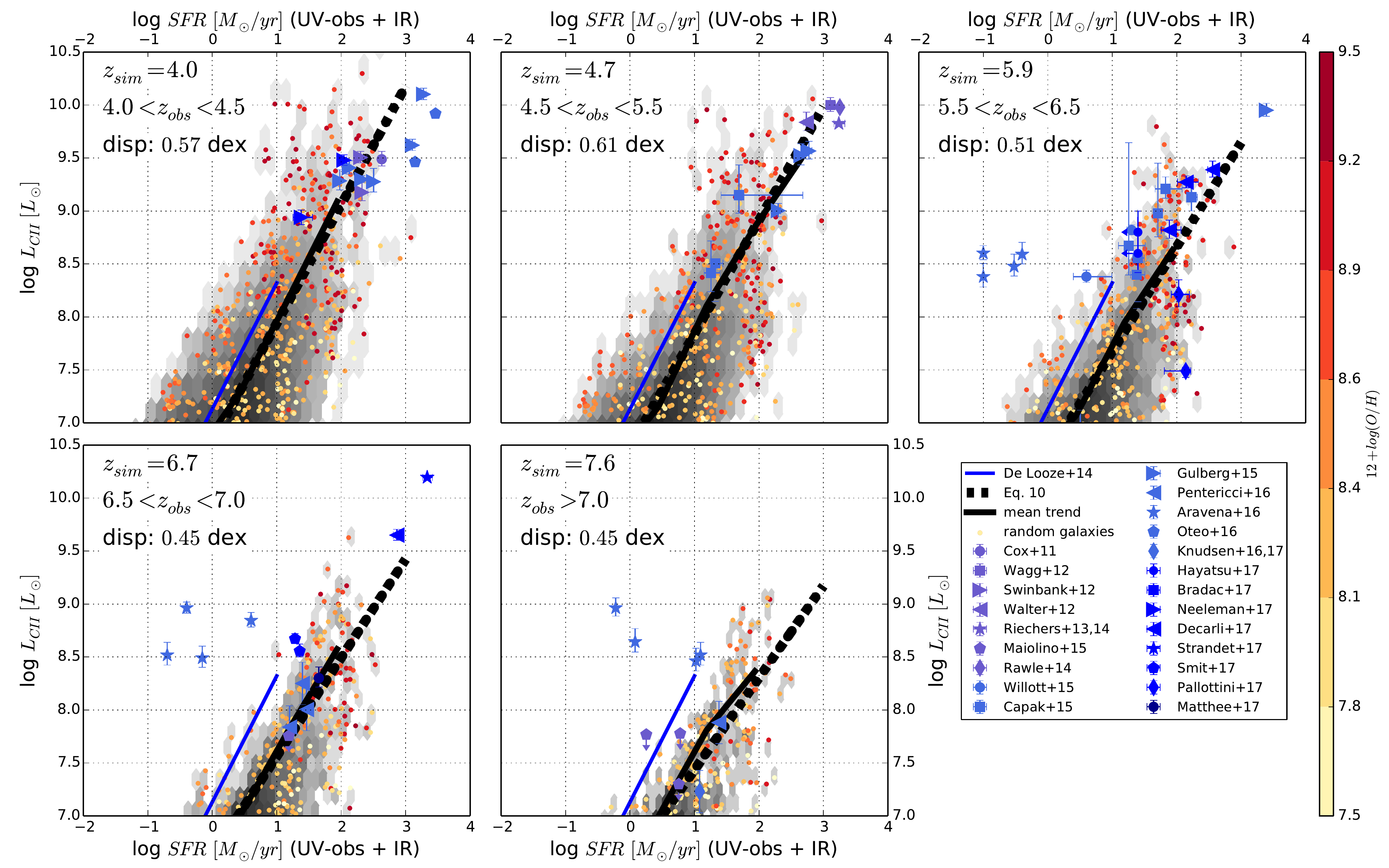}
\caption{L$_{\mathrm{[CII]}}$--SFR relation. Predictions from Lagache et al. (2018) model are shown for a set of redshifts from z = 4 to z = 7.6.  In each panel the whole sample of simulated galaxies is shown in grey scale. The average relation is plotted with a solid black line. Yellow to red coloured points mark the gas metallicity of a randomly selected sample of simulated galaxies (note that the observed tendency of high-metallicity galaxies to fall either above or below the mean trend, that is to make an envelope, is only a trick of the eye caused by the plotting; galaxies with high metallicities ($Z_g>8.8$) are spread over the whole area, with a higher density of objects at high SFR). 
The blue solid line shows the relation for the local dwarf galaxy sample (from De Looze et al. 2014).
The predictions from the model are compared to a large sample of observational data. {\it Figure extracted from Lagache et al. (2018)}. \label{SFR_CII}}
\end{center}
\end{figure}

\par\bigskip\noindent
\subsection{[CII] line intensity mapping}
[CII] is one of the brightest emission lines in the spectra of galaxies, and one of the most valuable tracers of dusty star formation at high redshift. The single [CII] fine structure transition is a very important coolant of the atomic interstellar medium (ISM) and of photon-dominated regions (PDRs). Carbon has an ionization potential of 11.3 eV (compared to 13.6 eV for hydrogen), implying that line emission can originate from a variety of phases of the ISM: cold atomic clouds (CNM), diffuse warm neutral and ionized medium, and HII regions. However, at high z, both observations (e.g., Stacey et al. 2010, Gullberg et al. 2015) and theoretical works (e.g., Vallini et al. 2015, Olsen et al. 2015, Pallottini et al. 2017a,b) show that [CII] originates mainly from the CNM and PDRs. Furthermore, at z$>$5-6, the CMB temperature approaches the spin temperature of the [CII] transition in the CNM suppressing the flux contrast (e.g., Lagache et al. 2018): hence, the [CII]
signal from high-z galaxies is dominated by the emission from PDR and is thus a unique tracer of SFR (see Fig.\,\ref{SFR_CII}). \\
So far, [CII] studies of very distant galaxies have been quite limited, with the detection of [CII] in $\sim$40 star-forming galaxies at $z>5$ (Lagache et al. 2018). No clear picture has emerged on the physical conditions of [CII] emission in these distant galaxies, but the ratio [CII] to LIR is surprisingly high at z$\sim$5 providing a hint that the environment of the interstellar medium conspires to keep [CII] bright (Capak et al. 2015). 
Moreover, [CII] is redshifted into the sub-millimeter and millimeter atmospheric windows for $4.5<z<8.5$ and is thus a line of choice for intensity mapping.

\begin{figure}[]
\begin{center}
\includegraphics[scale=0.24]{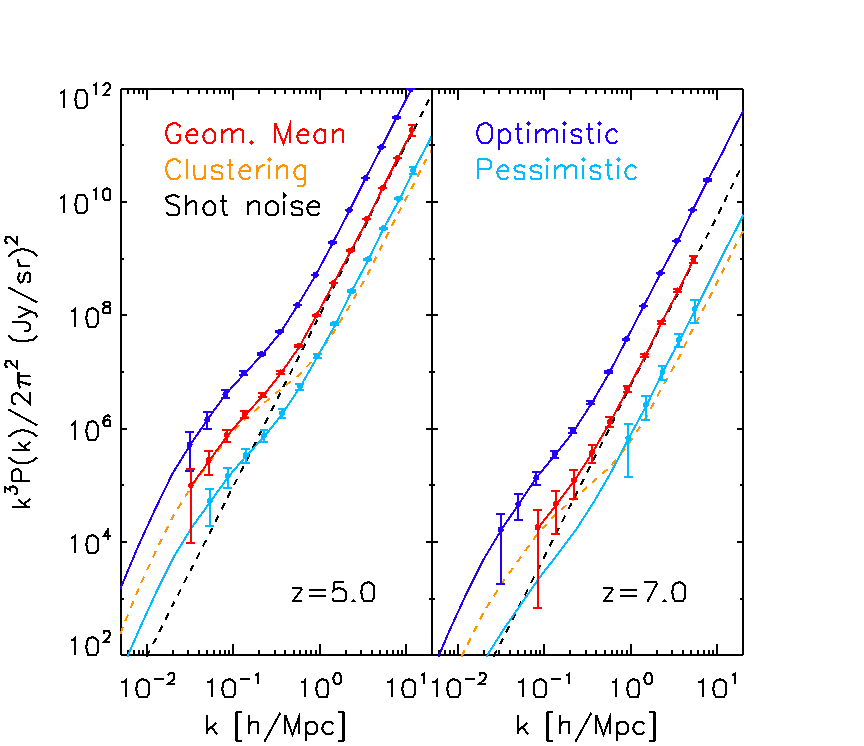}
\hspace{-1.cm}\includegraphics[scale=.23]{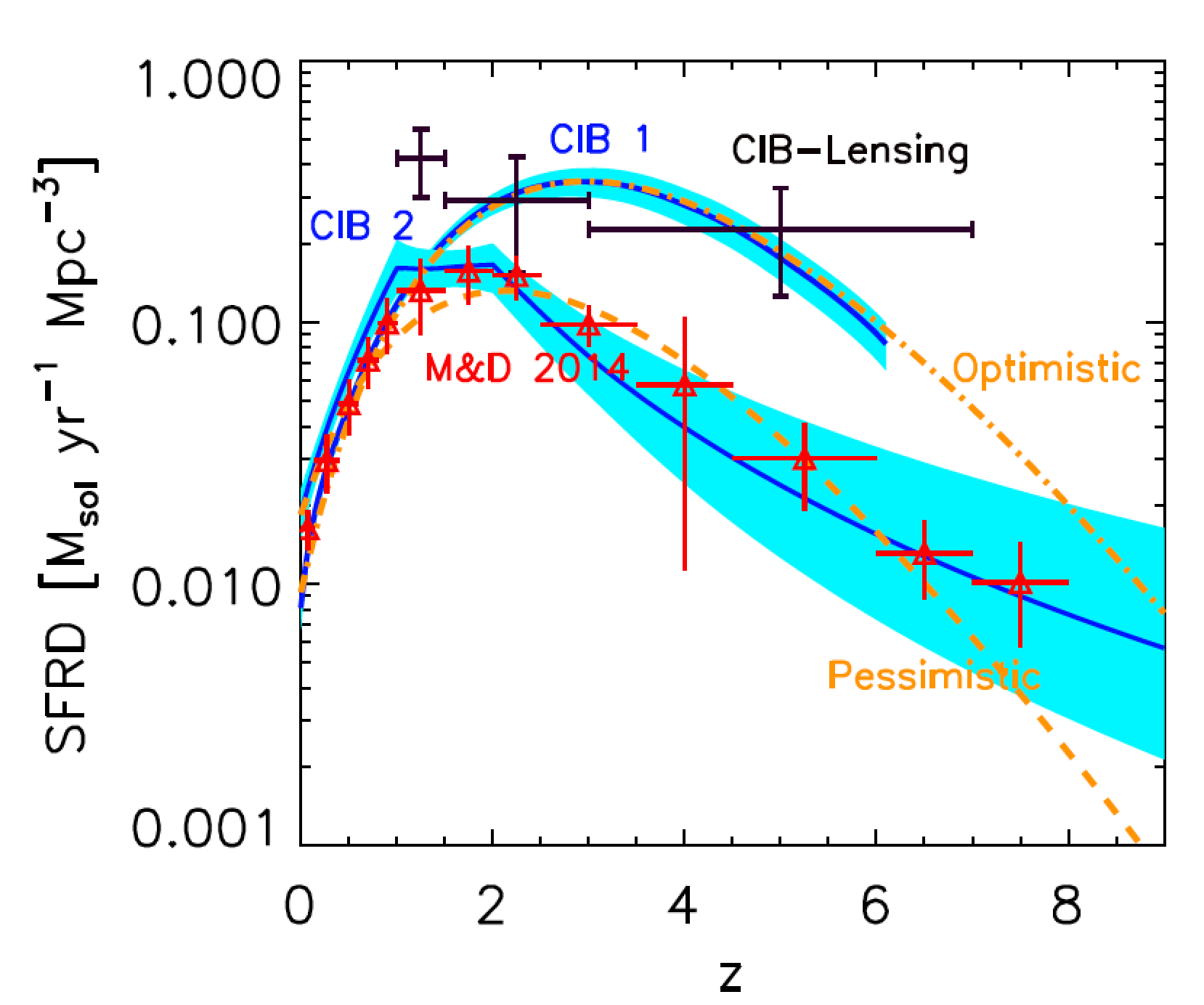}
\caption{{\it Left:} Predicted [CII] power spectrum (for three scenarios: optimistic, pessimistic and the geometrical mean) at redshifts z=5 and z=7. For the geometrical mean, the clustering and shot-noise terms are shown. Error bars have been
computed using the spectrometer characteristics, for a survey of 1.4 square degrees in 1,200 hours of observing time. Only points with SNR$>$1 are shown. {\it Right:} Figure illustrating how different the SFR densities derived from galaxies (Madau \& Dickinson 2014, red points) and CIB anisotropies are. Different models of CIB anisotropies, that fit equally well the measured power spectra, give very different SFR densities (cf CIB1 and CIB2 models: dark blue curves and 1$\sigma$ in light blue). Also shown are the measurements derived from the cross-correlation between the lensing map of the CMB and the CIB (black points, Planck collaboration 2014a). SFR densities derived from the two [CII] models we considered (optimistic and pessimistic cases) are shown in orange. At z$\gtrsim4$, the SFR density from the dusty star-formation is unknown (red points are exclusively coming from UV-selected galaxies). \label{concerto_fig} }
\end{center}
\end{figure}

\par\bigskip\noindent
\subsection{The CONCERTO project \label{concerto}}
We have designed a project called CONCERTO\footnote{https://people.lam.fr/lagache.guilaine/CONCERTO.html} (for CarbON CII line in post-rEionization and ReionizaTiOn epoch), whose goal is to construct and use a novel spectrometer to make pioneering measurements of the redshifted [CII] line to map the dusty star formation at high redshift and in the epoch of reionisation. CONCERTO instrument is based on the state-of-the-art development of new arrays in the millimeter using Kinetic Inductance Detectors (KIDs) and spectra will be obtained by a fast Martin-Puplett interferometer. The instrument is based on a dilution cryostat, not requiring liquid helium or nitrogen, and able to assure continuous operation, i.e. no recycling or other dead time. The instantaneous field of view (FOV) will exceed 150 arcmin$^2$ and the spectral resolution will be $\delta$=1.5\,GHz (corresponding to $\delta$z=0.04 at z=6). The instrument will cover the frequency band 200-360\,GHz (with a Ònotch filterÓ at 325\,GHz to suppress the background related to the water vapor line). The focal plane will be efficiently sampled with 1,500 pixels per array (one per linear polarization). The total number of pixels will thus be 3,000. It is planned to deploy CONCERTO to the APEX telescope, which is a 12-m antenna located at a 5105 m altitude on the Llano de Chajnantor in Northern Chile, providing an angular resolution of about 21 arcsec at 300\,GHz.\\

CONCERTO survey will provide a spatial-spectral data cube in which intensity is mapped as a function of sky position and redshift. The 3-D fluctuations are studied in Fourier space with the power spectrum.  We have used different models of intensity mapping of dusty star-forming galaxies (i.e. CIB anisotropies), and conversion factors between infrared luminosities and [CII] emission as observed in high-redshift galaxies (Lagache et al. 2018), to predict the [CII] power spectrum at $4.5<z<8.5$ (Fig.\,\ref{concerto_fig}). Because such predictions are very uncertain, we assumed two extreme models, giving respectively low and high SFR density at $z>3$. Low SFR density is the Pessimistic prediction as it corresponds to the lowest UV-driven SFRD; high SFR density is the Optimistic prediction as it corresponds to the CIB-driven SFR density. To compute the expected SNR, we have used realistic observational conditions and the instrument characteristics (sensitivities are computed using the NEFD already achieved with NIKA2 on sky, Adam et al. 2018). Covering an area of 1.4 square degrees, our survey will provide the first measurements of the [CII] power spectrum up to z$\sim$6 in the Pessimistic case, and up to z$\sim$8 in the Optimistic case. In the Pessimistic
case, the shot noise will be even detected up to z=7 (SNR=3 at k=2 h/Mpc). At z=8, while the [CII] power spectrum may not be detected, cross-correlation with other reionization probes may be more promising.

\section{Conclusion}
Intensity mapping is a promising approach for studying the dusty star-formation at high redshift. HI experiments will trace the reionization of the intergalactic medium, but they will not observe the young stars responsible for it. We thus propose as an alternative to map the star formation at redshift $4.5<z<8.5$ using intensity mapping of the [CII] 157.7 micron line. [CII] is one of the most valuable star formation tracers at high redshift.
With CONCERTO (see also the Time experiment, Crites et al. 2014), we will map the star formation at $z>4.5$, and deep into the epoch of reionization. We will use the [CII] line emission as a tracer of cosmic density structure, and give the first constraints on the power spectrum of dusty star-forming matter at these epochs. CONCERTO will also observe the CO intensity fluctuations arising from $0.3<z<2$ galaxies, giving the spatial distribution and abundance of molecular gas over a broad range of cosmic time.


\begin{thebibliography}{}
\bibitem[]{Adam2017} Adam, R., Adane A., Ade, P.A.R., et al. 2018, \textit{A\&A} in press, arXiv:1707.00908
\bibitem[]{bethermin2013}  B\'ethermin, M., Wang, L., Dor\'e, O., 2013, \textit{ApJ}, 557, 66
\bibitem[]{calura2017}  Calura F., Pozzi F., Cresci, G., et al. 2017, \textit{MNRAS}, 465, 54
\bibitem[]{capak2015} Capak, P. L., Carilli, C., Jones, G., et al. 2015, \textit{Nature}, 522, 455
\bibitem[]{crites2014} Crites A., Bock, J., J.,  Bradford, C.M., et al. 2014, \textit{SPIE},  9153, 1
\bibitem[]{delooze2014} De Looze, I., Cormier, D., Lebouteiller, V., et al. 2014, \textit{A\&A}, 568, A62
\bibitem[]{dole2006} Dole, H., Lagache, G., Puget, J.-L., et al. 2006, \textit{A\&A}, 451, 417
\bibitem[]{gispert2000} Gispert, R., Lagache, G.,  Puget, J. L. 2000, \textit{A\&A}, 360, 1
\bibitem[]{gullberg2015} Gullberg, B., De Breuck, C., Vieira, J. D., et al. 2015, \textit{MNRAS}, 449, 2883
\bibitem[]{kovetz2017} Kovetz, E.D., Viero, M.P., Lidz, A., 2018, submitted to Physics Reports, arXiv:1709.09066
\bibitem[]{lagache2005} Lagache, G., Puget, J.-L., Dole, H. 2005, \textit{ARA\&A}, 43, 727
\bibitem[]{lagache2018} Lagache, G., Cousin, M., Chatzikos, M., et al. 2018, \textit{A\&A} in press, arXiv:1711.00798
\bibitem[]{maniyar2018} Maniyar, A., Lagache, G., B\'ethermin, M., et al. 2018, accepted for publication in \textit{A\&A}
\bibitem[]{laporte2017} Laporte, N., Ellis, R. S., Boone, F., et al., 2017, \textit{ApJ}, 837, 21
\bibitem[]{madau2014} Madau, P. \& Dickinson, M.  2014, \textit{ARA\&A}, 52, 415
\bibitem[]{olsen2015} Olsen, K. P., Greve, T. R., Narayanan, D., et al. 2015, \textit{ApJ}, 814, 76
\bibitem[]{pallottini2017a} Pallottini, A., Ferrara, A., Gallerani, S., et al. 2017b, \textit{MNRAS}, 465, 2540
\bibitem[]{pallottini2017b} Pallottini, A., Ferrara, A., Bovino, S., et al. 2017a, \textit{MNRAS}, 471, 4128
\bibitem[]{planck2011}  Planck Collaboration XVIII. 2011, \textit{A\&A}, 536, A18
\bibitem[]{planck2014a} Planck Collaboration XVIII. 2014a, \textit{A\&A} 571, A18
\bibitem[]{planck2014b} Planck Collaboration XXX. 2014b, \textit{A\&A}, 571, A30
\bibitem[]{puget1996} Puget, J. L., Abergel, A., Bernard, J. P., et al. 1996, \textit{A\&A}, 768, 58
\bibitem[]{schmidt2015} Schmidt, S. J., MŽnard, B., Scranton, R., et al. 2015, \textit{MNRAS}, 446, 2696
\bibitem[]{stacey2010} Stacey, G. J., Hailey-Dunsheath, S., Ferkinho, C., et al. 2010, \textit{ApJ}, 724, 957
\bibitem[]{thacker2013} Thacker, C., Cooray, A., Smidt, J., et al. 2013, \textit{ApJ}, 779, 32
\bibitem[]{vallini2015} Vallini, L., Gallerani, S., Ferrara, A., Pallottini, A., Yue, B. 2015, \textit{ApJ}, 813, 36
\bibitem[]{viero2013} Viero, M. P., Moncelsi, L., Quadri, R. F., et al. 2013,  \textit{ApJ}, 779, 32
\bibitem[]{walter2012}  Walter, F., Decarli, R., Carilli, C., et al. 2012, \textit{Nature}, 486, 233
\bibitem[]{watson2015}  Watson, D., Christensen, L., Knudsen, K. K., et al. 2015, Nature, 519, 327

\end{thebibliography}
\end{document}